\documentstyle[twocolumn,psfig,floats,aps]{revtex}

\begin{document}
\twocolumn[\hsize\textwidth\columnwidth\hsize\csname
@twocolumnfalse\endcsname
\title{\bf Influence of the anion potential on the charge ordering 
in quasi-one dimensional charge transfer salts}
\author{Jos\'e~Riera$^{a}$ and Didier~Poilblanc$^{b,c}$}
\address{
$^a$Instituto de F\'{\i}sica Rosario, Consejo Nacional de
Investigaciones 
Cient\'{\i}ficas y T\'ecnicas, y Departamento de F\'{\i}sica,
Universidad Nacional de Rosario, Avenida Pellegrini 250, 2000-Rosario,
Argentina\\
$^b$Theoretische Physik, ETH-H\"onggerberg, CH-8093 Z\"urich, Switzerland\\
$^c$Laboratoire de Physique Quantique \& UMR--CNRS 5626,
Universit\'e Paul Sabatier, F-31062 Toulouse, France
}
\date{\today}
\maketitle

\begin{abstract}
We examine the various instabilities of quarter-filled strongly
correlated electronic chains in the presence of a coupling to the underlying
lattice. To mimic the physics of the (TMTTF)$_2$X Bechgaard-Fabre salts 
we also include electrostatic effects of intercalated anions.  
We show that small displacements of the anion can stabilize new mixed 
Charged Density
Wave-Bond Order Wave phases in which central symmetry centers
are suppressed. This finding is discussed in the context of 
recent experiments. We suggest that the recently observed charge
ordering is due to a cooperative effect between the Coulomb
interaction and the coupling of the electronic stacks to the 
anions. On the other hand, the Spin-Peierls instability at lower
temperature requires a Peierls-like lattice coupling.

\smallskip

\end{abstract}

\vskip2pc]



Site-centered Charge Density Wave (CDW) or Bond Order Waves (BOW)
states are generic in one-dimensional (1D) chains at commensurate
fillings. It is known that a strong enough short range electronic 
repulsion leads to a ($4k_F$) charge instability~\cite{review1,review2}
Molecular crystals such as quasi-1D
charge transfer salts~\cite{review1,review2} present a very rich physics
due to the interplay between electron-electron and electron-phonon 
interactions. Several systems have been observed to
show transitions towards charge ordered phases where the molecules
of the conducting stack exhibit unequal electron densities.
The sulfur series of the Bechgaard-Fabre salts family, 
namely $\rm (TMTTF)_2\, X$ ($\rm X=PF_6$, 
$\rm A\!sF_6$), consists of quarter-filled (${\bar n}=1/2$) 
molecular chains with very anisotropic
transport properties~\cite{review1}. Below the Mott 
localization crossover temperature~\cite{Pouget}
evidences for a true transition towards a symmetry broken
state have been recently
provided by dielectric response measurements~\cite{dielectric} and 
NMR~\cite{nmr_pf6}.
This lower symmetry phase is characterized by the disappearance of
all central symmetry centers in the crystal~\cite{dielectric}.
The Spin-Peierls (SP) transition observed at lower
temperature~\cite{review1} is characterized by a freezing of 
remaining spin fluctuations (opening of a spin gap) and 
it is accompanied by a tetramerization of the chains~\cite{review1}.

\begin{figure}[htb]
\vspace{-0.2truecm} 
\begin{center}
\psfig{figure=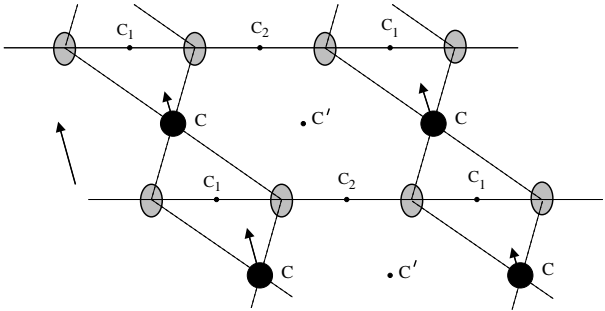,width=8truecm,angle=0}
\end{center}
\caption{
Schematic structure 
of the (a,b-c) plane (see Fig.~8 of Ref.~\protect\onlinecite{review1})
used in this paper: electronic chains (grey sites)
are separated by anionic chains (black sites) with one anion for two 
organic molecules. 
Due to the complete charge transfer of one electron per anion the
electronic chains are quarter-filled on average.
The arbitrary displacements (exaggerated for clarity) of the anions 
with respect to symmetric positions are shown by arrows. 
Central symmetry centers C, C', C$_1$ and C$_2$ in the high-symmetry 
configuration are indicated on the plot.
}
\label{lattice}
\end{figure}

The interplay between electronic correlations and
lattice effects in 1D quarter filled chains 
have been addressed in a number
of previous theoretical studies~\cite{Mazumdar1,Mazumdar2,diagad} where 
interesting modulated phases have been proposed. Generically,
lattice modulations (or BOW) were shown to be always accompanied by
CDW's of weaker amplitudes~\cite{diagad}.
However, the potential created by the intercalated anions 
have also been invoked~\cite{anions} as an extrinsic cause to the 
4k$_F$ dimerization. In that case, a weaker lattice coupling is
sufficient to produce a similar 2k$_F$ instability (tetramerization).  
In this work, we extend our previous model of Ref.~\onlinecite{diagad}
to include the crucial effect of the anion potential. The following
approach is based on the realistic two-dimensional (2D) 
structure of the charge transfer salts depicted in
Figure~\ref{lattice}
(for more details see also Figs.~1~\&~8 of Ref.~\onlinecite{review1} 
or Fig.~3.3 of Ref.~\onlinecite{review2}). 
We only focus here on the most anisotropic compounds of the sulfur 
series where inter-chain charge transfer can be safely neglected.

Our model consists of a 2D array of electronic 
(extended) Hubbard chains 
(hereafter called E-chains) coupled to classical lattice degrees of freedom;
\begin{eqnarray}
H&=& \sum_{i,j,\sigma} t(i,j) (c_{i-1,j;\sigma}^\dagger
c_{i,j;\sigma}+h.c.)+U\sum_{i,j} n_{i,j;\uparrow}n_{i,j;\downarrow}
\nonumber \\
&+& V\sum_{i,j} n_{i-1,j}n_{i,j} + H_{\rm elas} + H_{\rm anions}\, ,
\label{Ham}
\end{eqnarray}
where $i$ ($j$) are site (chain) labels,
$n_{i,j;\sigma}=c_{i,j;\sigma}^\dagger c_{i,j;\sigma}$ and 
$n_{i,j}=n_{i,j;\uparrow}+n_{i,j;\downarrow}$.
We have included a nearest neighbor (NN) interaction $V$.
Small local displacements of the molecules along the E-chains 
can couple strongly to the electrons via
modulations of the single particle hoppings of the Peierls type,
\begin{eqnarray}
t(2p-1,j)&=&t_1^0\, (1+\delta^B_{2p-1,j}),
\nonumber \\
t(2p,j)&=&t_2^0\, (1+g\delta^B_{2p,j}),
\end{eqnarray}
with an elastic energy 
\begin{equation}
H_{\rm elas}=\frac{1}{2}K_B \sum_{i,j} (\delta^B_{i,j})^2\, .
\end{equation}
The electron-lattice couplings have been (partially) absorbed in the 
re-definition of the bond modulations $\delta^B_{i,j}$
so that the strength of the lattice coupling scales like $1/K_B$.
Let us now consider the role of the charged $X^-$ anions 
located on the so-called A-chains as shown in Fig.~\ref{lattice}.
They introduce a new periodicity of 2 lattice spacings on
the E-chains which, a priori, might lead to a 
dimerization of the bare electronic transfer integrals along the chain 
i.e. $t^0_1\ne t^0_2$. Hereafter, the mean value $t=(t^0_1+t^0_2)/2$
is set to 1 and we assume $t^0_1-t^0_2=0.1 t$.
However, the qualitative conclusions of this work do not really depend
whether $t^0_1$ and $t^0_2$ are assumed to be equal or not.
For simplicity, we shall take $g=1$.
The anions can play a dominant role if they are allowed to undergo small
displacements (along some arbitrary direction) leading to 
local changes of the on-site electronic energies. In lowest order
in these displacements $\delta^A_{p,j}$ one gets,
\begin{eqnarray}
H_{\rm anions}=\sum_{p,j} \{ (\lambda\delta^A_{p,j}-\delta^A_{p,j+1})
n_{2p,j} \\
+ (\delta^A_{p,j}-\lambda\delta^A_{p,j+1}) n_{2p+1,j} \}
+\frac{1}{2}K_A \sum_{p,j}(\delta^A_{p,j})^2\, ,
\nonumber
\end{eqnarray}
where $p$ labels the positions of the anions in the chain direction.
Note that all displacements (for molecules and anions) are defined
with respect to the high-symmetry (high-T)
phase shown in Fig.~\ref{lattice}. The parameter $\lambda$ 
($0\le\lambda\le 1$) 
accounts for the non-equivalent location of each given anion with 
respect to the two nearest molecules in each of the two neighboring chains.
Hereafter, results are obtained for an intermediate value 
$\lambda=0.3$.

Following the method of Ref.~\onlinecite{diagad} 
based on Exact Diagonalisation (ED) of small clusters 
supplemented by a self-consistent procedure 
we obtain the lowest (T=0) energy lattice configuration without 
any assumption 
on the super-cell order of the broken symmetry ground state (GS).
The condition that the energy $E(\{\delta_{i,j}^B,\delta_{p,j'}^A\})$ 
is minimum with respect to the sets of distortions/displacements
$\{\delta_{i,j}^B\}$ and $\{\delta_{p,j'}^A\}$ reads,
\begin{eqnarray}
K_B\delta^B_{i,j} &+& t\big< c_{i-1,j;\sigma}^\dagger
c_{i,j;\sigma} + h.c.\big> =0 \label{non_linear}\, ,\nonumber\\
K_A\delta^A_{p,j'} &+& \lambda(\big<n_{2p,j'}\big>-\big<n_{2p+1,j'-1}\big>)\\
&+& \big<n_{2p+1,j'}\big>-\big<n_{2p,j'-1}\big> =0\, .
\nonumber
\end{eqnarray}
Here $\big<...\big>$ is the GS mean value obtained by ED (using
the Lanczos algorithm) of Hamiltonian (\ref{Ham}) on 
periodic $M\times L$ clusters. Typically, $M=2$ units ( 2 electronic
chains and 2 anionic chains) were
considered in the transverse
direction (so that modulated structures with $k_\perp=0$ or $\pi$ can 
spontaneously appear) with up to $L=12$ ($L/2=6$) electronic (anionic)
sites in the E-chains (A-chains).
The non-linear coupled equation~(\ref{non_linear}) can be solved by a
generalization of the iterative procedure of
Ref.~\onlinecite{Dobry_Riera}. Note that since spin and charge are 
conserved in each electronic chain, 
each of them can be diagonalized independently. 
Note also that, for finite lattice coupling, finite size effects 
are in general quite small. 

Our numerical results show a very rich phase diagram with mixed
CDW-BOW structures. Quite generally the largest supercell 
unit is four lattice spacings long along the E-chains
and has two basic (E-chain+A-chain) units in the transverse direction.
Therefore, in order to discuss our results, it is convenient to
parameterize the real space behavior of the various observables in the 
following way;
\begin{eqnarray}
\Delta n_{i,j}&=&\rho_{4k_F}(\!-1)^i\cos{(k_\perp j)}
\!+\!\rho_{2k_F}\cos{(\frac{\pi}{2}i\!+\!k_\perp j\!+\!\Phi_{2k_F})} \, , 
\nonumber\\
\delta_{i,j}^B&=&\delta^B_{4k_F}(\!-1)^i\cos{(k_\perp j)}
\!+\!\delta^B_{2k_F} \cos{(\frac{\pi}{2}i\!+\!k_\perp j\!+\!\Phi_{2k_F}^B)}
\, , \nonumber \\
\delta_{p,j}^A&=&\delta^A_{0} \cos{(k_\perp j)}+
\delta^A_{\pi}(\!-1)^p\cos{(k_\perp j)}
\, ,\label{parameterize}
\end{eqnarray}
where $\Delta n_{i,j}=(\big<n_{i,j}\big>-{\bar n})/{\bar n}$ is the
relative change of the electronic density. Note that the periodicity
of the 4k$_F$ (resp. 2k$_F$) modulation corresponds to 2 (resp. 4)
lattice spacing of the E-chains while a uniform ($k_\perp=0$) or 
a staggered ($k_\perp=\pi$) arrangement occur in the
transverse direction. Note that both the $4k_F$-BOW (dimerization),
the $2k_F$-BOW sequence X--0--${\bar{\rm X}}$--0 ($\Phi_{2k_F}^B=0$) and
the $2k_F$-CDW sequence B--B--${\bar{\rm B}}$--${\bar{\rm B}}$
($\Phi_{2k_F}=\pi/4$) preserve at least one of the central symmetries of
Fig.~\ref{lattice} (${\bar{\rm X}}=-X$).

Before discussing the generic phase diagram of the model, it is
interesting to comment on the special limit $K_A=\infty$, i.e 
corresponding to fixed anion positions ($\delta_{p,j}^A=0$). 
In that case, the E-chains decouple from each other and the
problem reduces to the single Peierls-Hubbard chain~\cite{Hirsch,diagad}.
The phase diagram of this model~\cite{diagad} exhibits, for sufficiently large 
lattice coupling, an instability towards two different 
translation symmetry breaking phases $D_1$ and $D_2$; 
(i) in the weak coupling regime (let's say $U/t<3$), a strong 
2k$_F$ BOW with $\Phi_{2k_F}^B=\pi/4$ 
i.e. corresponding to a X--X--${\bar{\rm X}}$-${\bar{\rm X}}$ 
type of sequence of the bonds
occurs.  This modulation coexists with a weaker 2k$_F$ site-centered 
CDW (A--0--${\bar{\rm A}}$--0 type of sequence of the 
on-site relative charge densities), 
and an even weaker 4k$_F$ (i.e. 
A--${\bar{\rm A}}$--A--${\bar{\rm A}}$) CDW component;
(ii) at larger $U/t$, a superposition of
a lattice dimerization (4k$_F$ BOW) together with a tetramerization
(2k$_F$ BOW with $\Phi_{2k_F}^B=0$) both 
coexisting with a weak $2k_F$ CDW component (with $\Phi_{2k_F}=\pi/4$)
occurs. It is interesting to notice that 
the dimerization
observed in experiments on Bechgaard-Fabre salts~\cite{review1}
is absent in the  weak coupling $D_1$ phase.
In addition, the $D_2$ phase, a natural candidate to describe the
low temperature SP phase, still exhibits one center of
symmetry (e.g. $C_2$) while recent experiments~\cite{dielectric} point
towards the disappearance of {\it all} center of symmetries 
below the charge ordering transition. 
Therefore, we believe that the anions may play an important role
in stabilizing new lower symmetry CDW-BOW phases. 

\begin{figure}[htb]
\vspace{-0.2truecm} 
\begin{center}
\psfig{figure=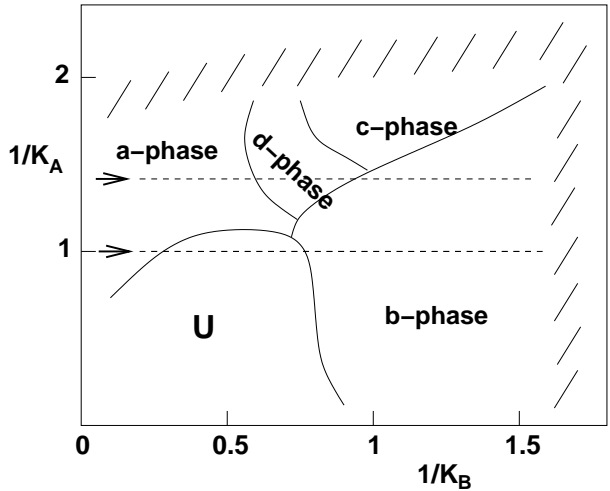,width=8truecm,angle=0}
\end{center}
\caption{Phase diagram versus the two electron-lattice couplings
for $U=6$, $V=2$ and $\lambda=0.3$. Dashed regions are beyond the
validity of the model. The different phases are depicted in
Fig.~\protect\ref{phases}. U denotes the uniform phase
(with only bare dimerization). 
The numerical values of the order parameters along the dashed lines
indicated by the arrows can be found in Fig.~\protect\ref{amplitudes}.
}
\label{phasediag}
\end{figure}

\begin{figure}[htb]
\vspace{-0.2truecm} 
\begin{center}
\psfig{figure=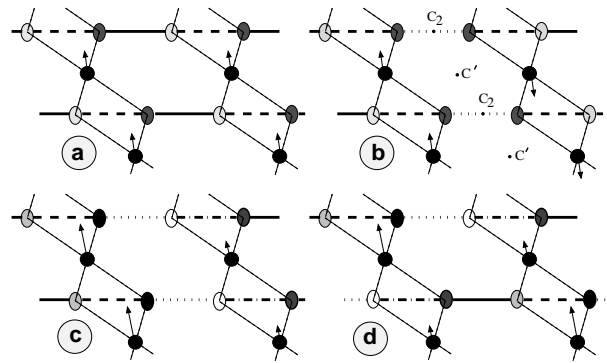,width=8truecm,angle=0}
\end{center}
\caption{Schematic representation of the various phases found in this
work. Different average charge densities are
indicated by different levels of grey. (a) and (b) [resp. (c) and (d)]
have two [resp. four] non-equivalent sites. While (a) and (b) 
show two and three non-equivalent bonds respectively, (c) and (d) have
four.
}
\label{phases}
\end{figure}

The T=0 phase diagram of our model is shown in  Fig.~\ref{phasediag}
for typical parameters in Bechgaard-Fabre salts. A rich variety of 
charge ordered phases are found and schematically depicted in
Figs.~\ref{phases}. We note that (at least on finite clusters) 
finite values of the electron-lattice couplings are necessary to
stabilize these translation symmetry breaking phases. 
At small Peierls coupling, the coupling to the anion displacement
field generates a 
${\bf k}_e=(4k_F,0)$ electronic CDW simultaneously
with a dimerization in the chain direction (hereafter 
named as "a-phase"). 
This charge ordering is accompanied by a uniform 
${\bf k}_A=(0,0)$ displacement 
of the anions with respect to their symmetric
positions  as shown in Fig.~\ref{phases}(a). 
On the other hand, the $D_2$ phase of the single chain
system (see Ref.~\cite{diagad}) at large enough $1/K_B$ coupling  
is weakly affected by the coupling to the
anions and continuously evolves into a two-dimensional 
in-phase array (i.e. $k_\perp=0$) 
of mixed $2k_F$-CDW/$2k_F$-BOW/$4k_F$-BOW chains 
(named as b-phase here) which 
extends in a wide region of the phase diagram. Note that, as soon as
$1/K_A$ is finite, this mixed CDW-BOW phase is
accompanied by a ${\bf k}_A=(\pi,0)$ component of the anion
displacement field (see Fig.~\ref{phases}(b)) so that
the translation symmetry is fully preserved in the transverse direction. 
As for the single chain $D_2$ phase, the two-dimensional b-phase 
arrangement preserves the symmetry centers $C_2$ and $C'$. 
Novel interesting phases are obtained 
in the region of intermediate $1/K_A$ and $1/K_B$ couplings.
The c-phase of Fig.~\ref{phases}(c) [resp. d-phase of  Fig.~\ref{phases}(d)]
is realized by a superposition
of a uniform ${\bf k}=(0,0)$ displacement and a ${\bf k}_A=(\pi,0)$ 
[resp. ${\bf k}_A=(\pi,\pi)$] modulation of the anion positions. 
The charge ordering is characterized by the superposition of
a $4k_F$-CDW with a $2k_F$-CDW whose phase $0<\Phi_{2k_F}<\pi/4$ 
depends on the model parameters. 
Similarly the lattice modulation contains a $4k_F$-component
(dimerization) superposed with an out-of-phase (i.e.
$0<\Phi_{2k_F}^B<\pi/4$) $2k_F$-BOW.
While all the E-chains are in-phase in the c-phase, they are
alternating in the d-phase. 
It is important to notice that only the uniform phase and the b-phase 
preserve at least one center of symmetry of the crystal.

\begin{figure}[htb]
\vspace{-0.2truecm} 
\begin{center}
\psfig{figure=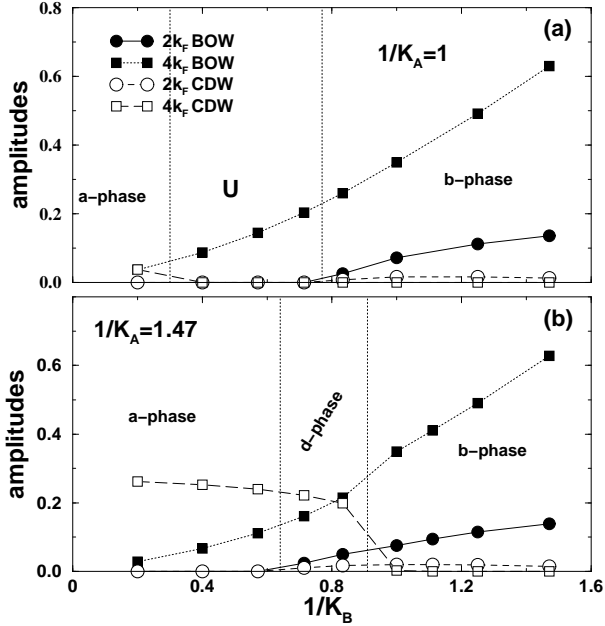,width=8truecm,angle=-90}
\end{center}
\caption{
Amplitudes of the various $2k_F$ and $4k_F$ CDW and BOW 
components (as indicated on the plot) as a function of the Peierls coupling 
$1/K_B$, for $U=6$, $V=2$, $\lambda=0.3$ and two different 
anion-electron coupling constants as indicated on the plots.
The correspondance to the ordered phases of Fig.~\protect\ref{phases}
is shown.}
\label{amplitudes}
\end{figure}

The modulation amplitudes are plotted vs $1/K_B$ in 
Figs.~\ref{amplitudes}(a,b) for two characteristic values of the 
electron-anion coupling constants. These plots show clearly that the
Peierls coupling stabilizes BOW's while the coupling to the anions
rather stabilizes $4k_F$-CDW.
We believe that the competition between these two effects
produces the interesting mixed 4-component CDW-BOW phases at intermediate
$1/K_A$ and $1/K_B$ values. 
In the c-phase and d-phase the phase of the $2k_F$ CDW [resp. $2k_F$
BOW] increases rapidly from $\sim 0$ to $\sim\pi/4$ [resp. decreases from 
$\sim\pi/4$ to $\sim 0$] for increasing Peierls coupling to 
eventually become locked to 
the value $\pi/4$ [resp. 0] in the b-phase (analogous to the 1D 
$D_2$ phase). Note that all center of symmetries are
removed by the $4k_F$-CDW component in the a-phase, c-phase and d-phase
and/or also by the  $2k_F$ CDW and/or BOW components
in the c-phase and d-phase as soon as 
$\Phi_{2k_F}\ne \pi/4$ and/or $\Phi_{2k_F}^B\ne 0$.
Note also that electronic interaction
(e.g. $U$) tends to suppress both $2k_F$ CDW and BOW components
as seen in Ref.~\cite{diagad}.

We finish by a discussion of our results in the light of the recent 
experiments~\cite{dielectric,nmr_pf6} providing evidences of a
charge modulated state in (TMTTF)$_2$X (X=PF$_6$, AsF$_6$). 
Although the spontaneously dimerized 
phase~\cite{diagad} ($D_2$ phase) of the (uniform) single chain 
Peierls-Hubbard model is a reasonable candidate for the charge ordered 
state~\cite{note}, it seems in contradiction with the conclusion of 
the dielectric response measurements due to its 
remaining central symmetry center.
Hence, although the single chain 
Peierls-Hubbard model contains most of the relevant features of the 
experimental system, 
a correct understanding still requires to consider the effective 
coupling of these (electronically decoupled) chains via the
anion potential. We believe that the dimerized a-phase stabilized
above a critical value of $1/K_A$ might be a fair description of
the ordered phase with 2 inequivalent sites in agreement
with NMR spectroscopy results~\cite{nmr_pf6}.
Clearly Coulomb interaction ($V$ term) also cooperates with the
electrostatic anion potential to generate the $4k_F$ CDW instability.
The small additional $2k_F$ BOW and CDW components of 
the c-phase (or d-phase) might also explain the additional
translation symmetry breaking (and the spin gap opening) occurring
at the SP transition. Indeed, although at $T=0$
we obtain first order lines between the various GS (by varying the
parameters), continuous transitions could appear at finite temperature
(and once lattice dynamics is introduced) between let say
the uniform phase and the a-phase at high temperature and the a-phase
and the c-phase (or d-phase) at lower temperature.
Incidently, this scenario then implies that the charge ordering 
must be accompanied by a (probably quite small) global shift of the
anions with respect to their symmetric locations
and that the subsequent SP transition is linked to
an additional (even smaller) ${\bf k}_A=(\pi,0)$ 
or ${\bf k}_A=(\pi,\pi)$ modulation.


Computations were performed at
the School of Computational Science \& Information
Technology (CSIT) and at the Academic Computing and Network 
Services at Tallahassee
(Florida) and at IDRIS, Orsay (France). 
Support from ECOS-SECyT A97E05 is also acknowledged.  
We are indebted to S.~Ravy for many useful suggestions and comments.

\end{document}